\documentclass[fleqn,usenatbib]{mnras}

\usepackage{graphicx}	
\usepackage{amsmath}	
\usepackage{amssymb}	
\usepackage{mathtools}
\usepackage{adjustbox}
\usepackage{threeparttable}
\usepackage{tabto}
\usepackage{float}
\usepackage{color}
\usepackage{graphics}
\usepackage{adjustbox}
\usepackage[usenames,dvipsnames]{pstricks}
\usepackage{epsfig}
\usepackage{pst-grad} 
\usepackage{pst-plot} 
\usepackage{times,pstricks,pst-node,pst-coil}
\usepackage{multicol} 
\usepackage{multirow} 
\usepackage{indentfirst} 
\usepackage{natbib} 
\usepackage{threeparttable}
\usepackage{longtable}
\usepackage{pdflscape}

\newcommand{\Msun}{M$_{\odot}$}

\definecolor{smalt(darkpowderblue)}{rgb}{0.0, 0.2, 0.6}
\definecolor{forestgreen(traditional)}{rgb}{0.0, 0.5, 0.0}

\newcommand{\vtau}{V471~Tau}

\newcommand{\bse}{{\sc bse}}
\newcommand{\mocca}{{\sc mocca}}

\newcommand{\chandra}{\emph{Chandra}}

\usepackage{etoolbox}

\title[Explaining the dearth of bright IPs in GCs]
{Magnetic dynamos in white dwarfs -- I. Explaining the dearth of bright intermediate polars in globular clusters}

\author[D. Belloni et al.]{
Diogo Belloni,$^{1,2}$\thanks{diogobellonizorzi@gmail.com (DB)}
Matthias R. Schreiber,$^{2,3}$\thanks{matthias.schreiber@usm.cl (MRS)}
Maurizio Salaris,$^{4}$
Thomas J. Maccarone$^{5}$
and
Monica Zorotovic$^{6}$
\\
$^{1}$ National Institute for Space Research, Av. dos Astronautas, 1758, 12227-010, S\~ao Jos\'e dos Campos, SP, Brazil\\
$^{2}$ Departamento de F\'isica, Universidad T\'ecnica Federico Santa Mar\'ia, Av. Espa\~na 1680, Valpara\'iso, Chile\\
$^{3}$ Millenium Nucleus for Planet Formation, Valpara{\'i}so, Chile\\
$^{4}$ Astrophysics Research Institute, Liverpool John Moores University, 146 Brownlow Hill, Liverpool L3 5RF, UK\\
$^{5}$ Department of Physics and Astronomy, Box 41051, Science Building, Texas Tech University, Lubbock, TX 79409-1051, USA\\
$^{6}$ Instituto de F{\'i}sica y Astronom{\'i}a, Universidad de Valpara{\'i}so, Av. Gran Breta{\~n}a 1111, 2360102, Valpara{\'i}so, Chile
}

\date{Accepted 2021 May 19. Received 2021 May 19; in original form 2021 April 29}

\pubyear{2021}

\begin{document}
\label{firstpage}
\pagerange{\pageref{firstpage}--\pageref{lastpage}}
\maketitle

\begin{abstract}
Recently, Bahramian et al. investigated a large sample of globular clusters (GCs) and found that bright intermediate polars (IPs) are a factor of $10$ less frequent in GCs than in the Galactic field.
We theoretically investigate here this discrepancy based on GC numerical simulations. 
We found that, due to disruptive dynamical interaction, there is on average a reduction of only half of bright IP progenitors, which is clearly not enough to explain the observed deficiency.
However, if the rotation- and crystallization-driven dynamo scenario recently proposed by Schreiber et al. is incorporated in the simulations, the observed rareness of bright IPs in GCs can be reproduced. 
This is because bright cataclysmic variables in GCs are typically very old systems ($\gtrsim10$~Gyr), with white dwarfs that almost fully crystallized before mass transfer started, which does not allow strong magnetic fields to be generated.
The observed mass density of bright IPs in GCs can be recovered if around one third of the bright cataclysmic variables dynamically formed through mergers have magnetic field strengths similar to those of IPs.
We conclude that the observed paucity of bright IPs in GCs is a natural consequence of the newly proposed rotation- and crystallization-driven dynamo scenario.
\end{abstract}


\section{Introduction}
\label{introduction}

\noindent
The study of globular clusters (GCs) plays an important role in our understanding of the Universe since they are natural laboratories for testing theories of stellar dynamics and evolution, very old
($\sim10-12$~Gyr) and have high central stellar densities (up to $\sim10^6$~stars per pc$^3$).
These properties make stellar dynamical interactions very frequent in these environments, which turns GCs into factories of exotic systems such as low-mass X-ray binaries, red and blue straggler stars, millisecond pulsars, and black hole binaries, among others.
One of the most abundant types of interacting binaries in GCs are cataclysmic variables (CVs), which are characterized by a white dwarf (WD) that stably accretes matter from a low-mass main-sequence (MS) star \citep[see][for comprehensive reviews on GC CVs]{Knigge_2012MMSAI,Belloni_Rivera_2021}.

Intermediate polars (IPs) are typically brighter (with X-ray luminosities $\sim10^{33}$~erg~s$^{-1}$ being fairly common for IPs with periods longer than $3$~hr and rare for other classes) and have harder spectra than other classes of CVs (again with $kT$ typically tens of keV for IPs, and only rarely so high for other classes).
Recently, \citet{Bahramian_2020} thoroughly investigated $38$~GCs with the \emph{Chandra} satellite, provided an extensive catalogue containing more than $1100$~X-ray sources in these GCs, 
and analysed in detail those sources with X-ray luminosities greater than $\sim10^{33}$~erg~s$^{-1}$, which includes bright IPs.
Based on the mass density of IPs in the Galactic disc, obtained from the space density \citep{Schwope_2018} and assuming a local stellar density of
$0.1$~pc$^{-3}$, \citet{Bahramian_2020} estimated that $16-34$ bright IPs should be present in their sample of GCs.
However, they found that only $2-3$ sources in those GCs are likely bright IPs.
This represents an under-abundance of a factor of $10$, at more than ${3\sigma}$ significance, in comparison with the Galactic field.
This goes in the opposite direction of previous observational works, which suggested that most GC CVs should be IPs \citep[e.g.][]{Grindlay_1995}.

There are currently four scenarios that have been suggested for the origin of magnetic fields in WD populations in the Galactic field.
In the \textit{fossil field scenario} the magnetic WD progenitor is most likely the product of a MS-MS binary merger. Such a merger can produce strong differential rotation and give rise to large-scale fields in the radiative envelopes of the merger outcome \citep{Ferrario2009,schneideretal19-1}.
In the \textit{common-envelope dynamo scenario} the magnetic field is generated when two stars merge during common-envelope evolution. In this case, the dynamo is expected to be driven by shear due to differential rotation either in the envelope  \citep[e.g.][]{potteretal10-1}, 
or in an accretion disc \citep{nordhausetal11-1}, or in the hot outer layers of the degenerate core \citep{wickramasingheetal14-1}.
In the \textit{double degenerate scenario}, a magnetic field is generated 
during the merger of two degenerate cores \citep{Berro_2012}.
More recently, \citet{Schreiber_2021} revised the model suggested by \citet{Isern_2017} and proposed that a \textit{rotation- and crystallization-driven dynamo} similar to those working in planets and low-mass stars can generate strong magnetic fields in the WDs in close binaries.

Utilizing the simulations carried out by \citet{Belloni_2019}, which corresponds to the largest set of GC models used to study CVs to date, we here investigate the possible reasons leading to the observed low incidence of bright IPs in GCs, taking into account the effect of dynamics in creating and destroying IP progenitors as well as the above-mentioned scenarios for the origin of magnetic WDs.

\vspace{-0.5cm}
\section{Globular cluster models}
\label{models}

\begin{figure}
  \begin{center}
     \includegraphics[width=0.95\linewidth]{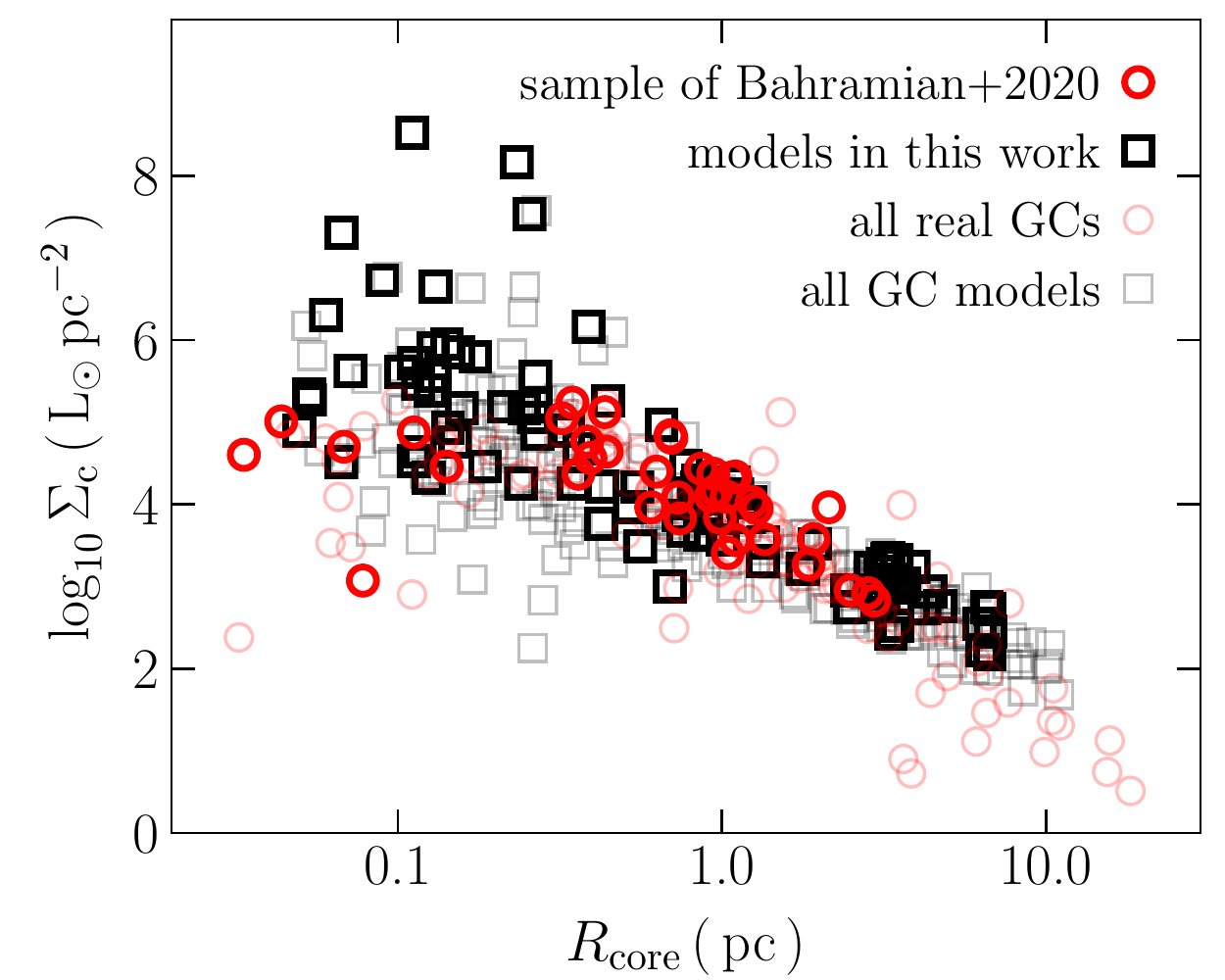}
  \end{center}
  \caption{Central surface brightness ($\Sigma_{\rm c}$) against the core radius ($R_{\rm core}$), for the \mocca~models (squares) and real GCs (circles), which observational data was extracted from \citet[][ updated 2010]{Harris_1996}. The figure illustrates that the models considered in this work are consistent with the sample of GCs analysed by \citet{Bahramian_2020}, except for just a few very bright compact models.}
  \label{Fig0}
\end{figure}

\noindent
In order to investigate bright IPs in GCs, we used the models simulated by \citet{Belloni_2019} with the \mocca~code \citep[][and references therein]{Hypki_2013}, which reproduces $N$-body direct simulations with great precision and accuracy \citep[e.g.][]{Madrid_2017}.
One very important part of \mocca~for this work is the Binary Star Evolution (\bse) code \citep{Hurley_2002}, which was properly upgraded and calibrated by \citet{Belloni_2018b} to deal with CV evolution.

Amongst our GC models, which have low metallicity (${Z=0.001}$) and assumed to be ${\approx12}$~Gyr old, we have a variety of different initial conditions spanning different values of the mass, the size, the King concentration parameter, the initial binary population, the binary fraction, and the Galactocentric distance.
In addition, we also explored two parameters of stellar/binary evolution, namely inclusion or not of mass fallback for black hole formation and three different common-envelope efficiencies.

We compare the distributions of the central surface brightness and the core radius of our models with those of observed GCs in Fig.~\ref{Fig0}.
These models are consistent with a substantial fraction of real GCs and can be considered as representative for the Galactic GC population.
In addition, models assuming the so-called Kroupa initial binary population \citep[][and references therein]{Belloni_2017c} and with low common-envelope efficiency (${\lesssim0.50}$) better reproduce the observed numbers of CVs and CV candidates in NGC~6397, NGC~6752, and 47~Tuc.
Taking this into account, from the $288$ GC models investigated in \citet{Belloni_2019}, we here only consider this sub-set, which correspond to $96$ GC models.
Fig~\ref{Fig0} also illustrates that, except for a very few bright compact models, this sub-set of models is consistent with the sample of GCs analyzed by \citet{Bahramian_2020}.

\vspace{-0.5cm}
\section{Bright Cataclysmic Variables in Globular Clusters}
\label{results}

\noindent
In \citet{Belloni_2019}, we investigated the impact of dynamics in creating CVs and destroying CV progenitors, taking into account the entire population of detectable CVs in the GC models.
We repeat here these analyses, but focusing only on \textit{bright CVs}, which are assumed here to be those located above the orbital period gap, since the X-ray bright IPs are dominantly in this region of the orbital period distribution \citep[e.g.][]{Suleimanov_2019}.
In particular, we investigate whether effective dynamical disruption of bright IP progenitors could explain the paucity of bright IPs in GCs.

\vspace{-0.5cm}
\subsection{Dynamical disruption of progenitors}

\noindent
\citet[][see their fig.~2]{Belloni_2019} found a strong correlation between the initial GC stellar encounter rate and the fraction of destroyed CV progenitors.
Since bright IPs are mostly concentrated above the orbital gap, these systems have to come from MS--MS binaries with relatively long initial orbital periods ($\sim10^3-10^5$~d).
This is because the pre-CV formation rate is much higher at the beginning of the GC evolution (within $\sim2$~Gyr after the GC formation) and only systems that start the CV phase close to the present day (i.e. spend $\sim10$~Gyr in the pre-CV phase) will likely be found above the orbital period gap. 
These systems have to leave the common-envelope evolution with longer orbital periods than those identified below the orbital period gap.
Their progenitors are therefore easier to be disrupted in dynamical interactions.

We found that the destruction fractions range from ${\sim10}$~per~cent (sparsest model) to ${\sim80}$~per~cent (densest model), and the average is ${\sim60}$~per~cent, which corresponds to an average reduction of a bit more than a factor of two.
Even though such a destruction rate of bright CV progenitors considerably reduces the incidence of these systems in GCs compared to the field, it is clearly not enough to explain the observed deficit of a factor of $10$ of bright IPs.
Therefore, \textit{the reason behind the observed low incidence of bright IPs in GCs must be related to the different characteristics of GC CVs, in comparison with their counterparts in the Galactic disc.}

\vspace{-0.5cm}
\subsection{Present-day population: formation channels and mass densities}

\noindent
As dynamical disruptive interactions of CV progenitors cannot explain the small number of observed bright IPs in GCs, we now take a look at the different formation channels of the predicted bright CVs with the aim of identifying which of these channels might perhaps not produce magnetic systems.
Of all bright CVs, common-envelope evolution of binaries undergoing no or very weak dynamical interactions is the dominant CV formation channel, producing on average $\sim89$~per~cent of bright CVs.
The remaining $\sim11$~per~cent were formed because of at least one strong dynamical interaction.

This predicted low average fraction of dynamically formed bright CVs can be compared with observational X-ray investigations considering the GC stellar encounter rates, which depends on the central density, the core radius and the central velocity dispersion, and can be used to address whether a particular type of X-ray source is predominantly created in GCs through dynamical interactions.
Our results seem to be in agreement with the \chandra~study performed by \citet{Cheng_2018}, who used a sample of 69 GCs, but to contradict previous \chandra~investigations, which considered much smaller GC samples \citep[e.g.][]{Pooley_2006}.
We restrict ourselves here to only mentioning that this sort of observational investigation should be taken with a grain of salt, since (i) the number of detected X-ray sources is very sensitive to the exposure time, (ii) these investigations are strongly X-ray biased and contamination from other types of sources are not rare, (iii) the properties of a particular GC might significantly change over its life-time, (iv) the size of the investigated sample does matter.
We can then conclude that even though such X-ray studies are helpful, follow-up researches involving  multi-wavelength observations, as homogeneous as possible, of a large sample of GCs are needed to fully investigate the dynamical origin of CVs in GCs.
A detailed discussion on this topic is provided in \citet[][their sections 2.4 and 4.2]{Belloni_Rivera_2021}.

Amongst the several dynamical formation channels for bright CVs in our simulations, the most prominent is three-body dynamical merger involving a single MS star and a MS--MS binary, leading to an MS--MS binary which later undergoes common-envelope evolution to become a WD-MS binary.
This formation channel corresponds to $\sim47$~per~cent of the dynamically formed CVs.
The second most efficient dynamical channel, corresponding to $\sim32$~per~cent of them, is three-body dynamical exchange in which a WD replaces the lighter MS star in an MS-MS binary, leading to a WD--MS binary.
The third most frequent channel involves mergers of WDs with MS stars in three-body or four-body interactions, which provides $\sim12$~per~cent of the dynamically formed CVs.
The last dynamical formation channel for bright CVs is either MS--MS or WD--MS merger in binary evolution, followed by dynamical exchange in three-body interactions, contributing to $\sim9$~per~cent of dynamically formed bright CVs.
The number of bright CVs, separated by their main formation channels, as well as their mass densities are summarized in Table~\ref{Tab}.

According to the available scenarios for the origin of magnetic WDs, we expect that a considerable portion of the bright CVs formed through mergers, which corresponds to ${\approx7}$~per~cent of the bright CVs, could be magnetic \citep[e.g.][]{Berro_2012,wickramasingheetal14-1,schneideretal19-1}.
The mass density of bright CVs formed through dynamical mergers is ${6\times10^{-7}}$~\Msun$^{-1}$, which is about three times larger than the observed mass density of bright IPs in GCs (${\sim2\times10^{-7}}$~\Msun$^{-1}$).
Since potentially a large number of these systems are magnetic, this formation channel alone can explain the amount of observed bright IPs in GCs.
On the other hand, mass densities of bright CVs not formed due to mergers is more than ${30}$ times larger than the observed mass density of bright IPs.
This implies that the bright CVs formed from either common-envelope evolution or exchange, which corresponds to ${\approx93}$~per~cent of the bright CVs, need to be virtually all non-magnetic in order to reproduce the observed rareness of bright IPs in GCs.

\vspace{-0.5cm}
\section{If not dynamics, what is causing the scarcity of bright intermediate polars in globular clusters?}

\noindent
We have shown in Section~\ref{results} that on average the vast majority of bright CVs in all our GC models form through common-envelope evolution and that virtually all of them need to be non-magnetic in order to explain the observed dearth of bright IPs in GCs.
Considering that the rotation- and crystallization-driven dynamo scenario is the only currently available scenario that can explain the characteristics of magnetic WDs in close binaries in the Galactic field, we investigate in what follows the implications of this model for the $\approx93$~per~cent of GC bright CVs that were not formed through dynamical mergers.
In other words, we will verify whether this scenario is also able to explain the characteristics of these systems in GCs, which implies that the model needs to predict that nearly all bright CVs not formed through dynamical mergers are non-magnetic.
We start with a brief review of the recently proposed dynamo scenario.

\begin{table}
\begin{center}
\caption{Relative frequency of bright CVs averaged over all our $96$~GC models, separated by their formation channels. In the last column we provide their mass density, having in mind that the observed mass density of bright IPs in GCs is $\sim2\times10^{-7}$~\Msun$^{-1}$.}
\setlength\tabcolsep{6pt} 
\renewcommand{\arraystretch}{1.4} 
\begin{tabular}{lrrc}
\hline
formation channel & number & fraction & mass density~~(\Msun$^{-1}$) \\
\hline
common envelope & $272$ &  $88.9$\,\%  & $7.10\times10^{-6}$ \\
exchange        &  $11$ &   $3.6$\,\%  & $2.87\times10^{-7}$ \\
merger          &  $23$ &   $7.5$\,\%  & $6.00\times10^{-7}$ \\
all             & $306$ & $100.0$\,\%  & $7.99\times10^{-6}$ \\
\hline
\label{Tab}
\end{tabular}
\end{center}
\end{table}

\vspace{-0.5cm}
\subsection{The rotation- and crystallization-driven dynamo}

\noindent
Crystallization of carbon-oxygen WDs leads to the formation of an oxygen rich solid core surrounded by a convective carbon rich liquid. As first noticed by \citet{Isern_2017}, this configuration a solid core embedded in a convective liquid could generate strong magnetic fields if the WD is rotating rapidly in full analogy to the mechanism responsible for generating the magnetic fields of planets and low-mass stars \citep{christensenetal10-1}.

\citet{Schreiber_2021} extended the work of \citet{Isern_2017} by incorporating the dynamo in evolutionary tracks of CVs taking into account the WD spin period and orbital period evolution. The WD spin period is usually very long after the WD formation, but can become very short, near break-up spin period, when the WD is accreting mass and angular momentum, as in CVs.

The resulting scenario for the formation and evolution of magnetic CVs  proposed by \citet{Schreiber_2021} can be summarized as follows.
The WDs in close post-common-envelope binaries are born non-magnetic.
Due to angular momentum losses, this non-magnetic detached binary evolves towards shorter orbital periods, while the WD cools and its core may eventually starts to crystallize.
When the secondary fills its Roche lobe, the onset of mass transfer occurs and the binary becomes a CV.
Since the WD accretes mass and angular momentum, it spins up.
If the WD core is crystallizing, the conditions for the rotation- and crystallization-driven dynamo are satisfied and the magnetic field is generated.
Only if the WD is already substantially crystallized before mass transfer starts or if it is still too young for the onset of crystallization, the conditions are not mutually met and the CV will remain non-magnetic.

As described in detail in \citet[][]{Schreiber_2021}, the new scenario can solve several long-standing problems in the Galactic field at once, namely the absence of magnetic WDs amongst young close detached binaries \citep{Parsons_2021}, the higher/lower incidence of magnetism in CVs/detached binaries than in single WDs \citep{Kepler_2013,Pala_2020,Parsons_2021}, the existence of the radio pulsar AR\,Sco \citep{marshetal16-1}, and the relative frequency of magnetic and non-magnetic CVs \citep{Pala_2020}.
Given that the new scenario is successful for close detached WD binaries in the Galactic field, in what follows, we test whether this scenario can also explain the dearth of bright IPs in GCs.

\vspace{-0.25cm}
\subsection{Applying the rotation- and crystallization-driven dynamo scenario to cataclysmic variables in globular clusters}

\noindent
In the context of the rotation- and crystallization-driven dynamo scenario, two major differences between GCs and the Galactic disc need to be considered.
First, while GCs have measured ages of ${\sim11-12}$~Gyr, the Galactic disc is younger (${\sim10}$~Gyr old).
The second is related to the different history of star formation.
In GCs, we expect one or more bursts of star formation within the first 
$10^8$~years of the cluster life \citep[e.g.][]{Bastian_2018}.
On the other hand, in the Milky Way disc, assuming a nearly uniform star formation rate until the present time reasonably well explains several observational constraints \citep[e.g.][]{Schulz_2015}.
Taking into account the different ages and star formation rates, we should expect somewhat different CV populations in the Galactic disc and in GCs. 
In particular, CVs in GCs should be much older than those in the Galactic disc \citep[see also][]{Belloni_2017a}.

\begin{figure}
  \begin{center}
     \includegraphics[width=0.95\linewidth]{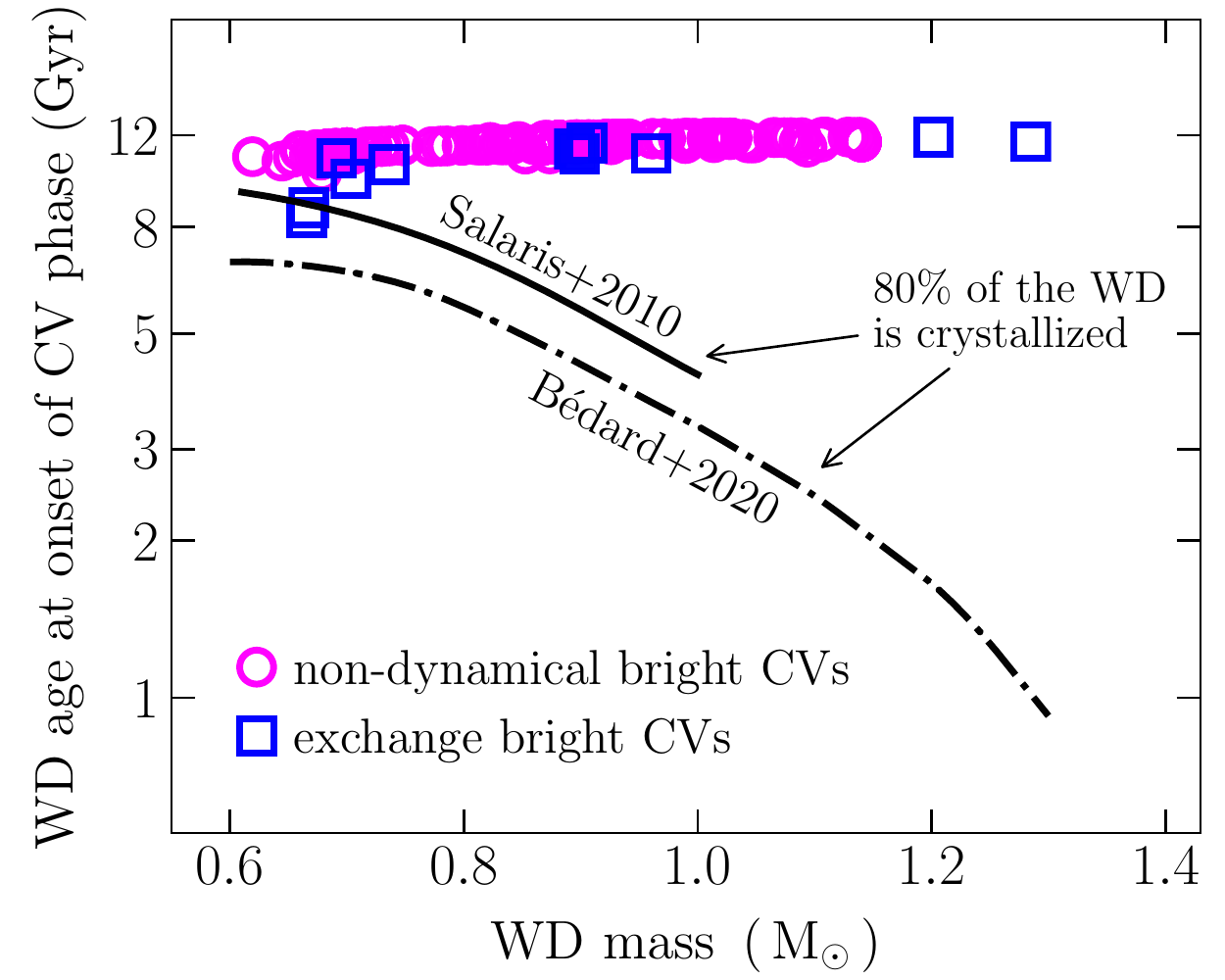}
  \end{center}
  \caption{WD ages at onset of CV phase against their masses for bright CVs formed either without strong dynamical interactions (magenta circles) or due to dynamical exchanges (blue squares). The solid and dot-dashed lines indicate when $80$~per~cent of the WD matter has crystallized, according to the models computed by \citet{Salaris_2010} and \citet{Bedard_2020}, respectively. Strong magnetic field generation due to the rotation- and crystallization-driven dynamo is expected for those systems lying well below these lines. Clearly, the new scenario does not predict any magnetic CV in this population.}
  \label{Fig}
\end{figure}

We show in Fig.~\ref{Fig} the WD age at onset of mass transfer for bright CVs formed either from isolated binaries or due to dynamical exchanges and compare it with the age at which $80$~per~cent of the WD matter has crystallized, according to the hydrogen-atmosphere WD models computed by \citet{Salaris_2010} and \citet{Bedard_2020}.
The differences between the ages in both works are mainly due to the different physics inputs and carbon-oxygen chemical stratifications, and the fact that the energy release due to phase separation upon crystallization is not included in \citet{Bedard_2020} calculations.

According to the model proposed by \citet{Schreiber_2021}, a sufficiently strong magnetic field due to a rotation- and crystallization-driven dynamo is expected for those systems having partially crystallized WDs, with crystallized mass fractions below ${\sim80}$~per~cent.
Fig.~\ref{Fig} clearly illustrates that all bright CVs formed without involving mergers are too old to develop any strong magnetic field.
Despite the fact that accretion heating during nova cycles could in principle slowly melt a fraction of the crystallized core, this process is expected to take several hundred Myr \citep[e.g.][]{Epelstain_2007}.
Given that this time-scale is longer than that of CV evolution above the orbital period gap, we expect a 
negligible number of magnetic bright CVs (if any) assuming the fields are generated by the rotation- and crystallization-driven dynamo scenario.
Taking into account that these channels correspond to $\approx93$~per~cent of all bright CVs, the scenario for the origin of magnetic WDs in close binaries predicts a significant under-abundance of bright IPs in GCs.

Therefore, this new scenario for WD magnetic field generation naturally solves the issue investigated here, and even independently on the impact of dynamics in GC modelling, since even for initially sparse clusters, in which dynamics contribute to only ${\sim10}$~per~cent of reduction, the new scenario would still guarantee that the vast majority of the population of bright CVs would be non-magnetic.
This implies that 
\textit{the rotation- and crytallization-driven dynamo scenario is not only able to explain the characteristics of magnetic CVs in the Galactic disc, but also of those in GCs}.

\vspace{-0.5cm}
\section{Bright intermediate polars from mergers}
\label{resultsMERGER}

\noindent
So far, we have shown that no bright IP is expected in the population of bright CVs formed from either primordial binary evolution or through dynamical exchanges if the new dynamo model is correct.
Therefore,
\textit{we expect that the observed population of bright IPs in GCs should be all dynamically formed through either MS--MS mergers or mergers involving WDs.}

In case of an MS--MS merger, \citet{schneideretal19-1} showed that strong, large-scale surface magnetic fields can be produced in coalescing MS binaries.
In cases of mergers involving WDs, mostly WD--MS and WD--WD mergers, a dynamo can be produced due to differential rotation in the outer layers of the merger product, resulting in strong magnetic fields \citep[e.g.][]{wickramasingheetal14-1,Berro_2012}.
However, we shall emphasize that it is not clear under which condition the magnetic field strengths generated in mergers would be in the range of those observed amongst IPs (${\sim1-10}$~MG), since the fields could be either stronger or weaker, which would not make them as bright in X-rays as IPs.
For instance, a correlation between the differential rotation and the field strength is expected in mergers involving WDs \citep[e.g.][]{wickramasingheetal14-1}.
While objects differentially rotating near break-up velocities are expected to develop the strongest fields, only weak fields should be generated with less differential rotation.
The situation is similar for MS--MS mergers. It is not clear how strong the magnetic field generated in MS--MS mergers could be in the resulting WD.

An example for the formation of weakly magnetized WDs in close binaries most likely through MS--MS merger is \vtau.
This post-common-envelope binary hosts a K-type MS star paired with a magnetic WD, which is the hottest, youngest and most massive WD in the Hyades open cluster.
In order to explain this clear evolutionary contradiction, \citet{OBrien_2001} proposed that such a WD is descended from a blue straggler, with mass about twice the turnoff mass, which was the product of a MS--MS merger.
When this star evolved and became a red giant, it filled its Roche lobe and the binary underwent common-envelope evolution, which led the binary to shrink to its present orbital period. 
The weak magnetic field of the WD in \vtau~clearly excludes this system to become a bright IP. 
This example illustrates that, despite mergers likely lead to magnetic field generation, many of the emerging magnetic WDs may not become IPs. 
Therefore, the predicted mass density of bright CVs formed through mergers (${6\times10^{-7}}$~\Msun$^{-1}$) should be treated as an upper limit for the predicted mass density of bright IPs.

Also the observed mass density of bright IPs in GCs of ${\sim2\times10^{-7}}$~\Msun$^{-1}$ \citep{Bahramian_2020} should be taken with a grain of salt.
These authors only analysed $38$~GCs, which is a relatively small sample of the entire Galactic GC population, and it has been already shown that results from X-ray investigations can strongly depend on the size of the sample \citep{Cheng_2018}.
In addition, they only provide bright IP candidates in their catalogue, lacking spectroscopic confirmation (or, at least, properly characterization of these candidates, e.g. by obtaining orbital periods), which might change the estimated mass density upon confirmation or not of some of these candidates.
Taking into account these uncertainties and given that the mass density estimated from observation is only a factor of three below the upper limit predicted by our simulations, we conclude that model predictions and observations are in very good agreement. 
The observed mass density could be explained, provided that roughly one third of the bright CVs formed through mergers have field strengths consistent with IPs, which appears reasonable.

\vspace{-0.8cm}
\section{Summary and Conclusions}

\noindent
We investigated bright CVs and IPs formed through either isolated binary evolution or dynamical exchange/merger interactions, in 96 GC models evolved with the \mocca~code with the aim of explaining the observed under-abundance (of a factor of $10$) of bright IPs in GCs compared to the Galactic disc.

We showed that dynamical disruption of bright IP progenitors in GCs is not enough to explain the observed dearth of bright IPs, since dynamics only provide on average a reduction of a bit more than a factor of two.
Many progenitor binaries survive in these dense environments and successfully become CVs, in addition to other CVs that are formed because of dynamical exchange.
These two groups correspond to ${\approx93}$~per~cent of all bright CVs and belong to a predominantly old population, in which their WDs are almost fully crystallized before the onset of mass transfer.
According to the recently proposed rotation- and crystallization-driven dynamo scenario, these characteristics do not allow strong magnetic field generation, which then explain the observed scarcity of bright IPs in GCs.

Finally, the remaining ${\approx7}$~per~cent of bright CVs were dynamically formed through mergers and likely harbour magnetic WDs.
The mass density of this group is about three times larger than the observed mass density of bright IPs.
Thus, if about one third of such CVs have magnetic field strengths comparable to those observed in IPs, the observed mass density of bright IPs can be recovered.

We conclude that the newly proposed rotation- and crystallization-driven dynamo scenario for the generation of magnetic fields in close white dwarf binaries naturally explains the observed paucity of bright IPs in GCs.
Additionally, the few bright IPs currently observed should be dynamically formed through mergers, involving either WDs or main-sequence stars that later become WDs.

\section*{Acknowledgements}

\noindent
We would like to thank an anonymous referee for the comments and suggestions that helped to improve this manuscript.
DB was supported by ESO/Gobierno de Chile and by the grant \#{2017/14289-3}, S\~ao Paulo Research Foundation (FAPESP).
MRS acknowledges financial support from FONDECYT grant number 1181404. 
MZ acknowledges support from CONICYT PAI (Concurso Nacional de Inserci\'on en la Academia 2017, Folio 79170121) and CONICYT/FONDECYT (Programa de Iniciaci\'on, Folio 11170559).

\vspace{-0.45cm}
\section*{Data availability}

\noindent
The data underlying this article can be obtained upon request.

\vspace{-0.55cm}
\bibliographystyle{mnras}
\bibliography{references}


\bsp	
\label{lastpage}

\end{document}